\documentclass{ws-ijmpcs}

\usepackage{graphicx}
\usepackage[caption=false]{subfig}

\usepackage{pifont}
\usepackage{color}
\usepackage{hyperref}

\newcommand{\pip}{\ensuremath{\pi^{+}}}
\newcommand{\pim}{\ensuremath{\pi^{-}}}
\newcommand{\pizero}{\ensuremath{\pi^{0}}}
\newcommand{\SigmaPlus}{\ensuremath{\Sigma^{+}}}
\newcommand{\SigmaZero}{\ensuremath{\Sigma^{0}}}
\newcommand{\SigmaMinus}{\ensuremath{\Sigma^{-}}}
\newcommand{\proton}{\ensuremath{p}}
\newcommand{\kp}{\ensuremath{K^{+}}}
\newcommand{\Flatte}{Flatt\'{e}}
\newcommand{\gevcc}{\ensuremath{\mathrm{GeV}/c^{2}}} % GeV/c2
\newcommand{\costhetakp}{\ensuremath{\cos \theta_{\kp}^{\mathrm{c.m.}}}}

\begin{document}

%\markboth{Moriya Schumacher}
%{Excited Hyperon Measurements at CLAS}

\catchline{}{}{}{}{}
%
%%%%%%%%%%%%%%%%%%%%%%%%%%%%%%%%%%%%%%%%%%%%%%%%%%%%%%%%%%%%%%%%%%%%

\title{Measurement of Excited Hyperons in Photoproduction at CLAS}

\author{Kei Moriya}

\address{Department of Physics, Indiana University, \\  % 727 E. Third St.,
Bloomington, IN 47405-7105,
USA\\
kmoriya@indiana.edu}

\author{Reinhard A. Schumacher}

\address{Department of Physics, Carnegie Mellon University, \\
Pittsburgh, PA 15213, USA\\
schumacher@cmu.edu}

\author{for the CLAS Collaboration}

\maketitle

\begin{history}
\received{\today}
\revised{\today}
\end{history}

\begin{abstract}
  Measurement results of photoproduced excited hyperon states
  using the CLAS detector at Jefferson Lab are shown. The invariant mass
  distribution of the $\Lambda(1405)$ has recently been shown to
  be different for each of the three $\Sigma \pi$ channels that it decays
  to, showing that there is prominent interference between
  the isospin $I=0$ and $I=1$ isospin amplitudes. Measurements of the
  differential and total cross sections of the three hyperons
  $\Lambda(1405), \Sigma^{0}(1385)$, and $\Lambda(1520)$ are presented and
  compared. Prospects of future studies using a $12$~GeV beam with the
  GlueX detector are briefly given.
 
\keywords{hyperons; $\Lambda(1405)$; $\Sigma(1385)$; $\Lambda(1520)$; CLAS;
  GlueX; photoproduction.}
\end{abstract}

\ccode{PACS numbers: 11.25.Hf, 123.1K}

\section{Introduction}
Photoproduction of excited hyperon states has not yet been studied
in detail compared to the production of the ground state hyperons. It
has been suggested that some of the so-called ``missing'' $N^{\ast}$
resonances may have relatively strong couplings to states besides $\pi
N$, for example the $K^{+} Y$ states, which may lead to a solution to
the missing baryon problem. Recent works~\cite{Anisovich} show that
indeed some of the resonances have couplings to the
ground state $K^{+} \Lambda$ and $K^{+} \Sigma^{0}$
states.

While the coupling to intermediate $N^{\ast}$ states is an interesting
problem, the nature of the excited hyperon states themselves is also
an interesting topic, as are the photoproduction mechanisms. Below we
present some of the recent excited hyperon measurements on the
$\Lambda(1405), \Sigma^{0}(1385)$, and $\Lambda(1520)$.

\section{$\Lambda(1405)$ Line Shape Measurements}

The $\Lambda(1405)$ is a state of interest due to its
peculiar $\Sigma \pi$ invariant mass spectrum, or ``line shape''. It
has been suggested in chiral unitary coupled channel studies that the
$\Lambda(1405)$ is a dynamically generated
resonance~\cite{1405-theory}, and a calculation within this framework
predicted that each of the different $\Sigma \pi$ line shapes would be
different~\cite{Nacher}. In recent years, there have been several
measurements of the $\Lambda(1405)$~\cite{1405-exp} in various
reactions, and this has lead to an exciting interplay between theory
and experiment. Together with the $\Sigma(1385)$ and $\Lambda(1520)$,
these are the lightest hyperon resonances, with spin-parity
numbers of $1/2^{-}, 3/2^{+}$, and $3/2^{-}$, respectively.
As each hyperon has different quantum numbers, a
comparison of the production of the three states may reveal
characteristics that are specific to the state or are otherwise common
to the photoproduction mechanisms of each state.

\begin{figure}[pb]
 \subfloat{\label{fig:fits:W2} \includegraphics[width=0.495\textwidth]{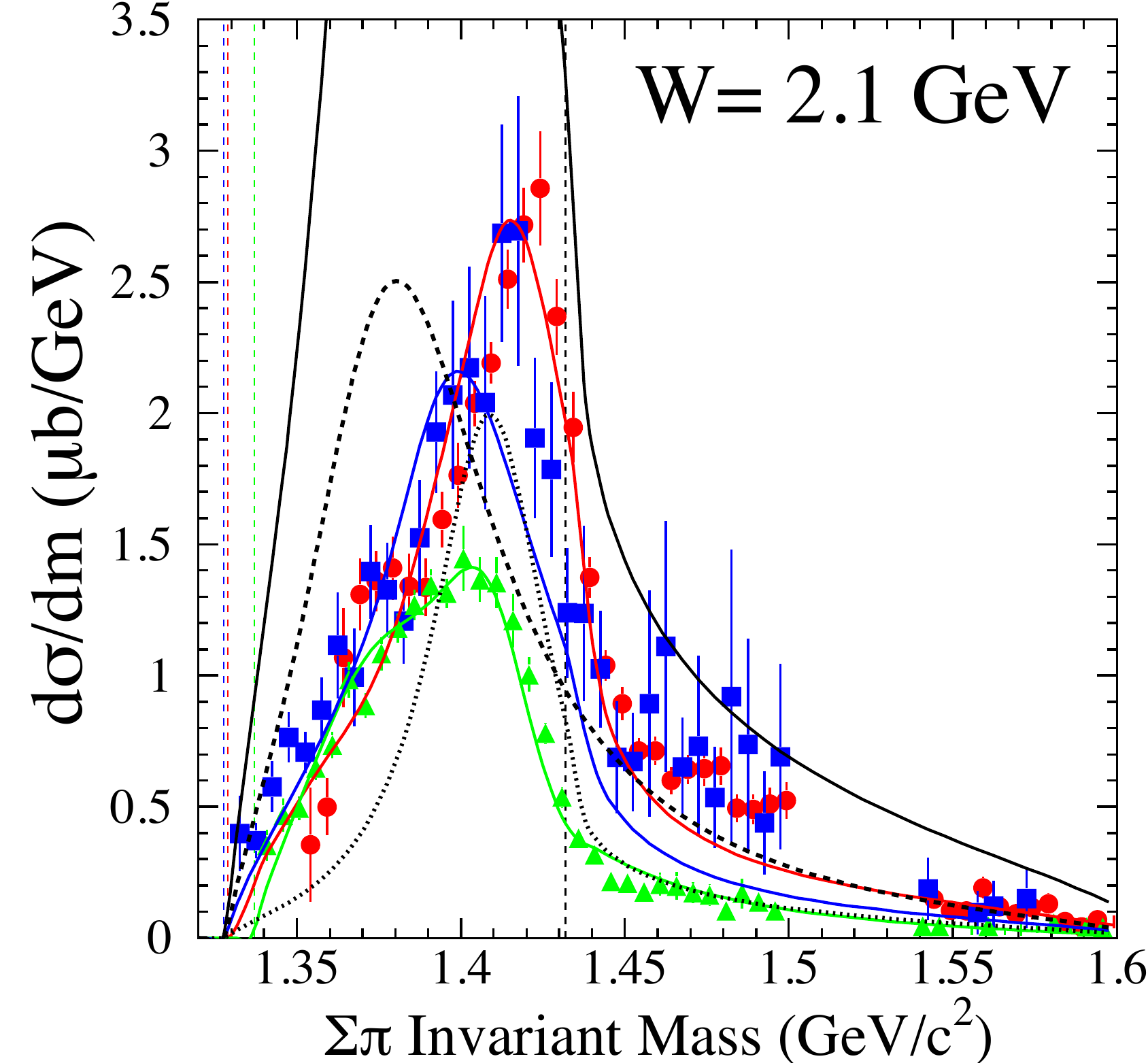}}
 \subfloat{\label{fig:fits:W8} \includegraphics[width=0.495\textwidth]{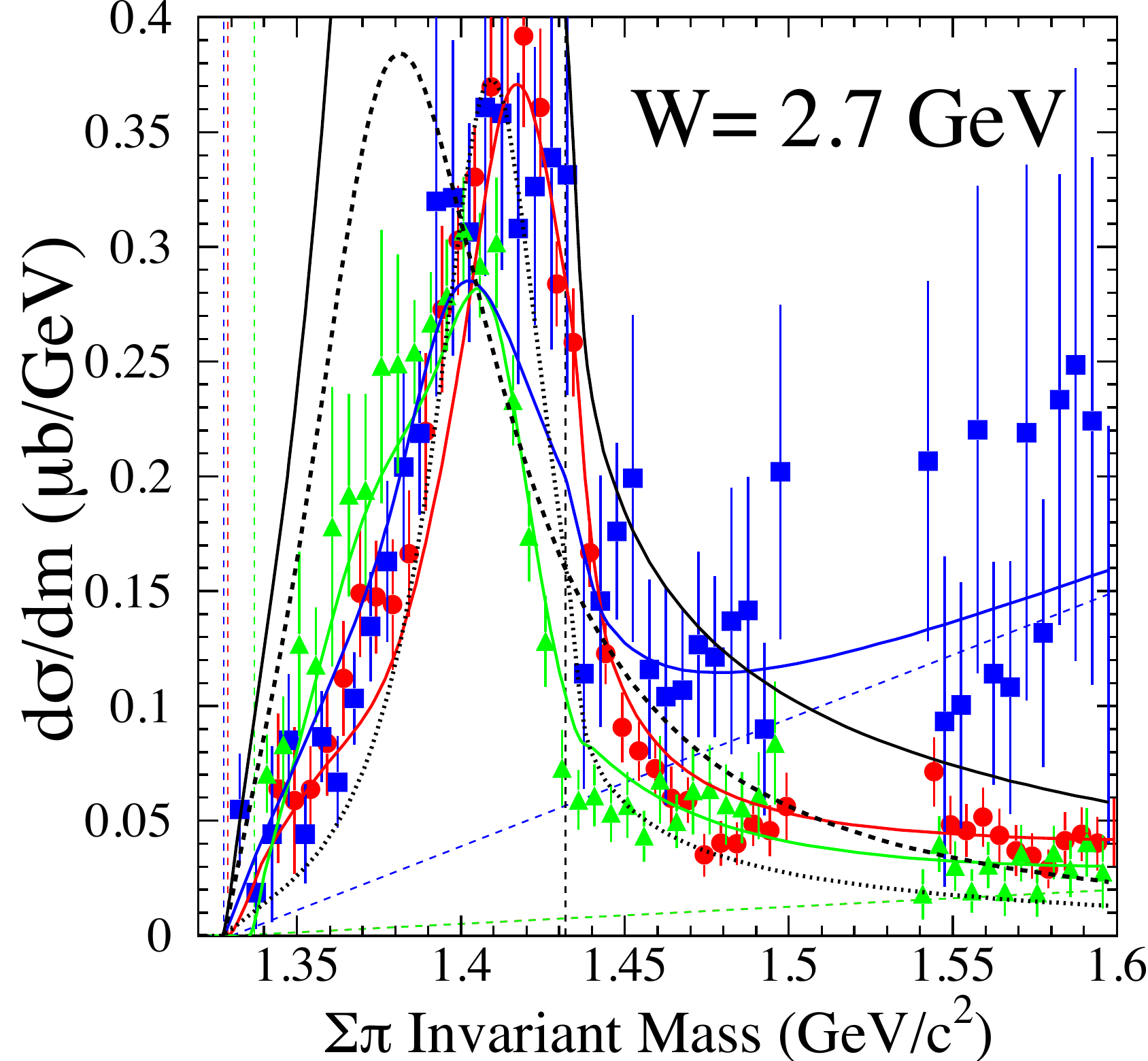}}
\vspace*{8pt}
\caption{Line shape of the $\Lambda(1405)$ for each $\Sigma
  \pi$ channel in two bins of $W$. The symbols are
  $\SigmaPlus \pim$ (red circles), $\SigmaZero \pizero$ (blue
  squares), $\SigmaMinus \pip$ (green triangles). The data are from
  Ref.~\protect\refcite{lineshape}, and
  the curves from Ref.~\protect\refcite{hyp12}. The dashed vertical
  line shows the $N \overline{K}$ threshold, where a drop in the
  intensity is seen for all line shapes.
\label{fig:fit}}
\end{figure}

The CLAS detector was located in Hall B of the Thomas Jefferson
National Accelerator Facility (JLab), and utilized a real photon beam
for photoproduction
experiments. We report on the reactions $\gamma + \proton \to K^{+}
Y^{\ast}$, where $Y^{\ast}$ is one of
$\Lambda(1405)$, $\Sigma^{0}(1385)$, or $\Lambda(1520)$. The
subsequent decays are into the $\Sigma \pi$ channels for the
$\Lambda(1405)$ and $\Lambda(1520)$, and $\Lambda \pizero$ for the
$\Sigma^{0}(1385)$. Details of the analysis can be found in
Refs.~\refcite{lineshape} and \refcite{xsec}. The line shapes of the
$\Lambda(1405)$ in two
center-of-mass (\mbox{c.m.}) energy bins $W$ are shown in
Fig.~\ref{fig:fit}.

Fits to extract the pole positions were performed
in Ref.~\refcite{hyp12}, and examples of the fit results are
shown as the curves Fig.~\ref{fig:fit}. The fits were done with relativistic
Breit-Wigner functions that had a \Flatte{}-type
modification~\cite{Flatte} to accommodate the strong coupling to the $N
\overline{K}$ state above that threshold, and assigned an isospin of
either $I=0$ (solid and dashed black) or $1$ (dotted black) so that
the contributions from pure isospin amplitudes
could be accounted for in each $\Sigma \pi$ channel. The amplitudes were
allowed to interfere coherently with each other, and reproduced the data
well\cite{Roca-fits}.
While measurements of all three $\Sigma \pi$ line shapes
represents great progress in the data from photoproduction, more
data from other reactions are wanted since theory predicts that the
population of the two poles of the $\Lambda(1405)$ depends on the
reaction~\cite{1405-theory}.

\section{Cross Sections}

\begin{figure}[h!t!pb]
 \subfloat{\label{fig:xsec_1405:W2} \includegraphics[width=0.49\textwidth]{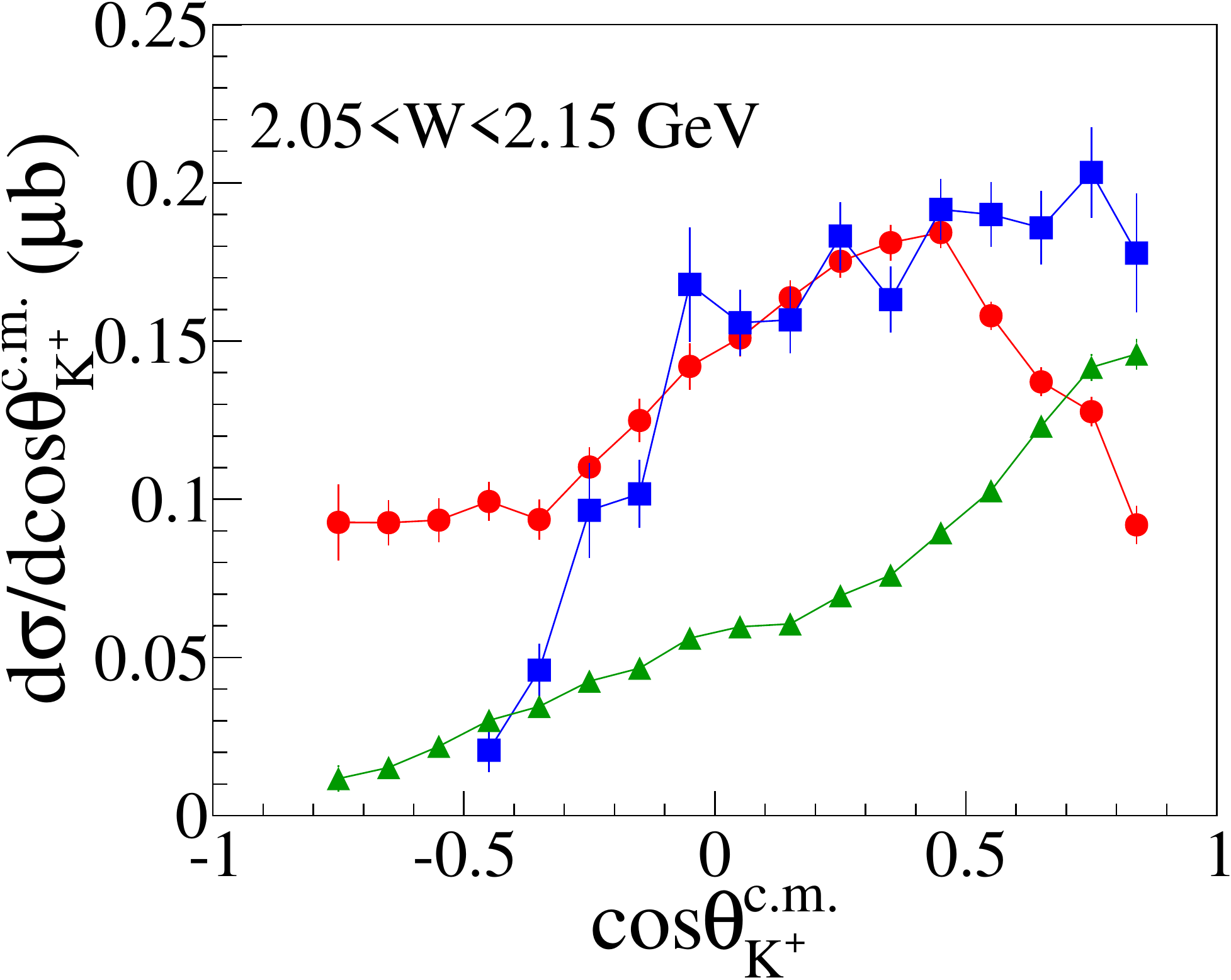}}
 \subfloat{\label{fig:xsec_1405:W8}\includegraphics[width=0.49\textwidth]{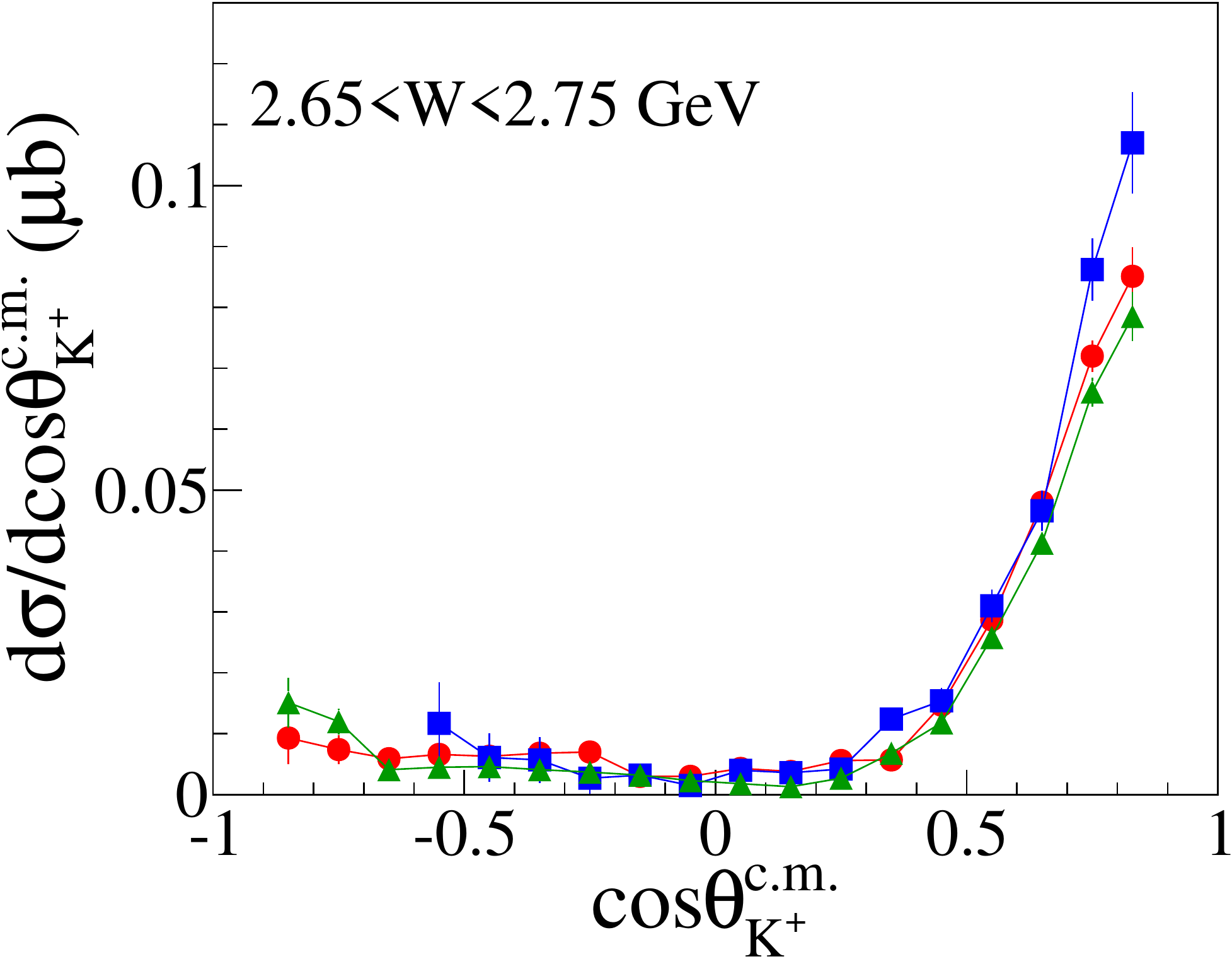}}
\vspace*{8pt}
\caption{Comparison of the differential cross sections for each $\Sigma
  \pi$ channel of the $\Lambda(1405)$. The data points for $\SigmaPlus
  \pim$ are shown with red circles, $\SigmaZero \pizero$ with blue
  squares, and $\SigmaMinus \pip$ with green triangles.
\label{fig:xsec_1405}}
\end{figure}

Measurements of the photoproduction differential cross sections of the
$\Lambda(1405), \Sigma^{0}(1385)$, and $\Lambda(1520)$ were recently
reported in Ref.~\refcite{xsec}. We begin with the
differential cross sections of the three $\Sigma \pi$ channels from
the $\Lambda(1405)$. Due to the presence of higher mass states, the
range of integration was fixed to be from $\Sigma \pi$ threshold up to
a $\Sigma \pi$ mass of $1.5$ \gevcc. Fig.~\ref{fig:xsec_1405}
compares the three $\Sigma \pi$ channels for two $W$
bins, and are plotted against \costhetakp.
Fig.~\ref{fig:xsec_1405}(\subref*{fig:xsec_1405:W2}) is near
production threshold, and we see stark qualitative differences between
the three channels. This result was totally unanticipated, and further
strengthens the case that there is a strong interference among
different isospin channels near the $\Lambda(1405)$. If this is the
case, this shows that the isospin interference is dependent not only
on the \mbox{c.m.} energy, but also on the production angle of the
$\Lambda(1405)$.
It is interesting to note that this difference appears only in the
near-production-threshold bins, where the difference in line shapes is
prominent. Figure~\ref{fig:xsec_1405}(\subref*{fig:xsec_1405:W8})
is for a higher
energy bin, and here we see that all three channels exhibit the same
kind of forward-peaked behavior.

\begin{figure}[h!t!pb]
 \subfloat{\label{fig:xsec_cmp:W2} \includegraphics[width=0.49\textwidth]{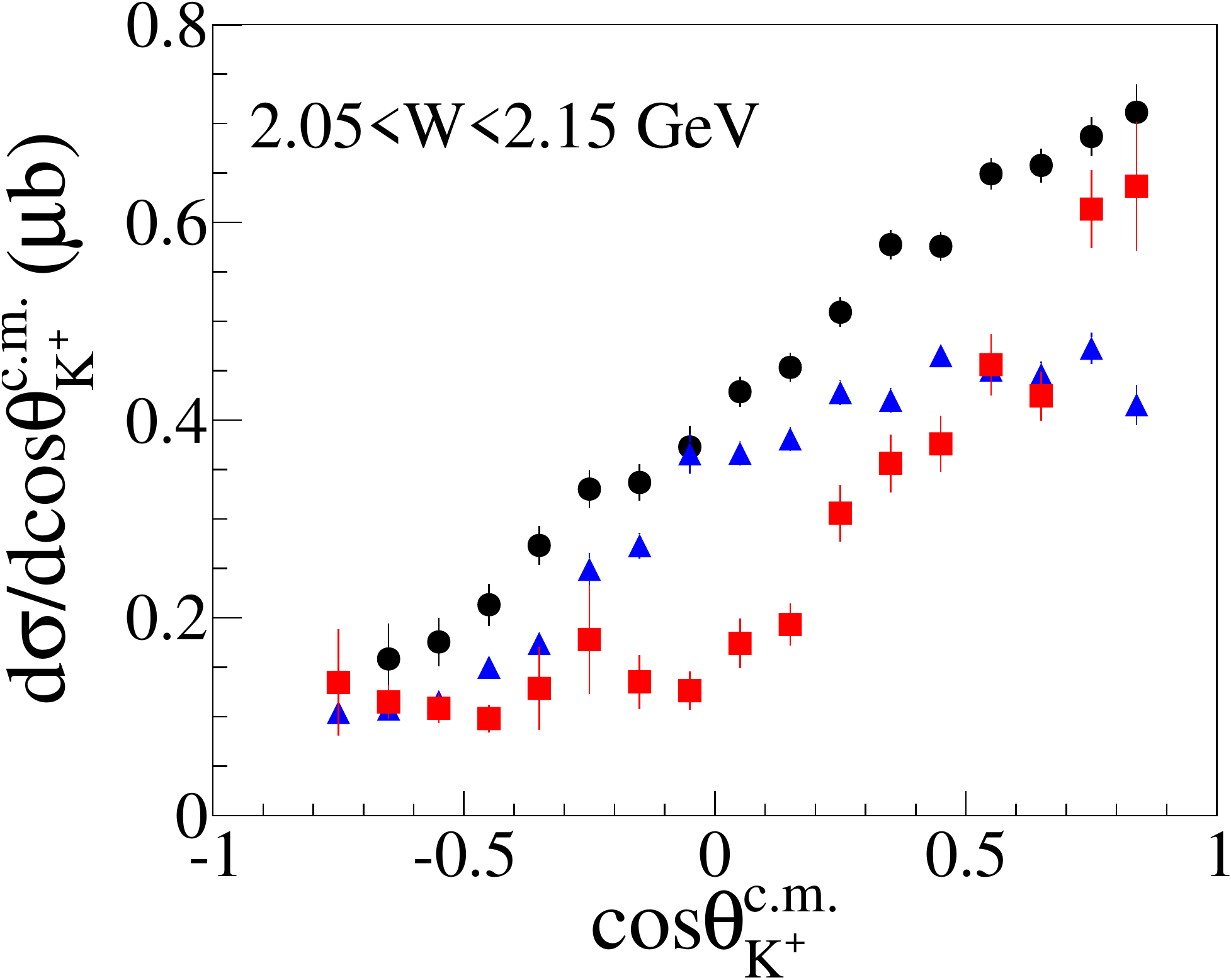}}
 \subfloat{\label{fig:xsec_cmp:W4} \includegraphics[width=0.49\textwidth]{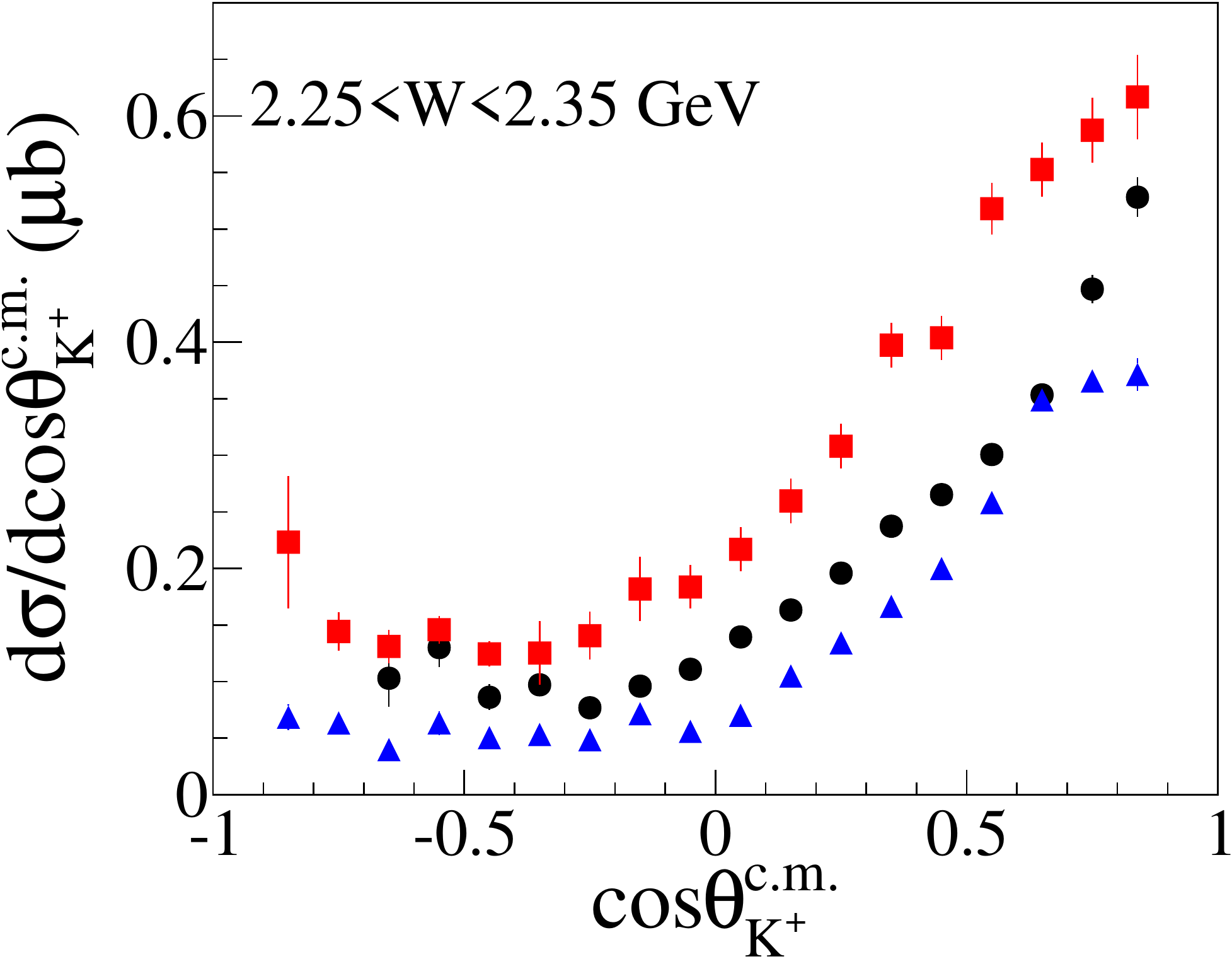}} \\
 \subfloat{\label{fig:xsec_cmp:W6} \includegraphics[width=0.49\textwidth]{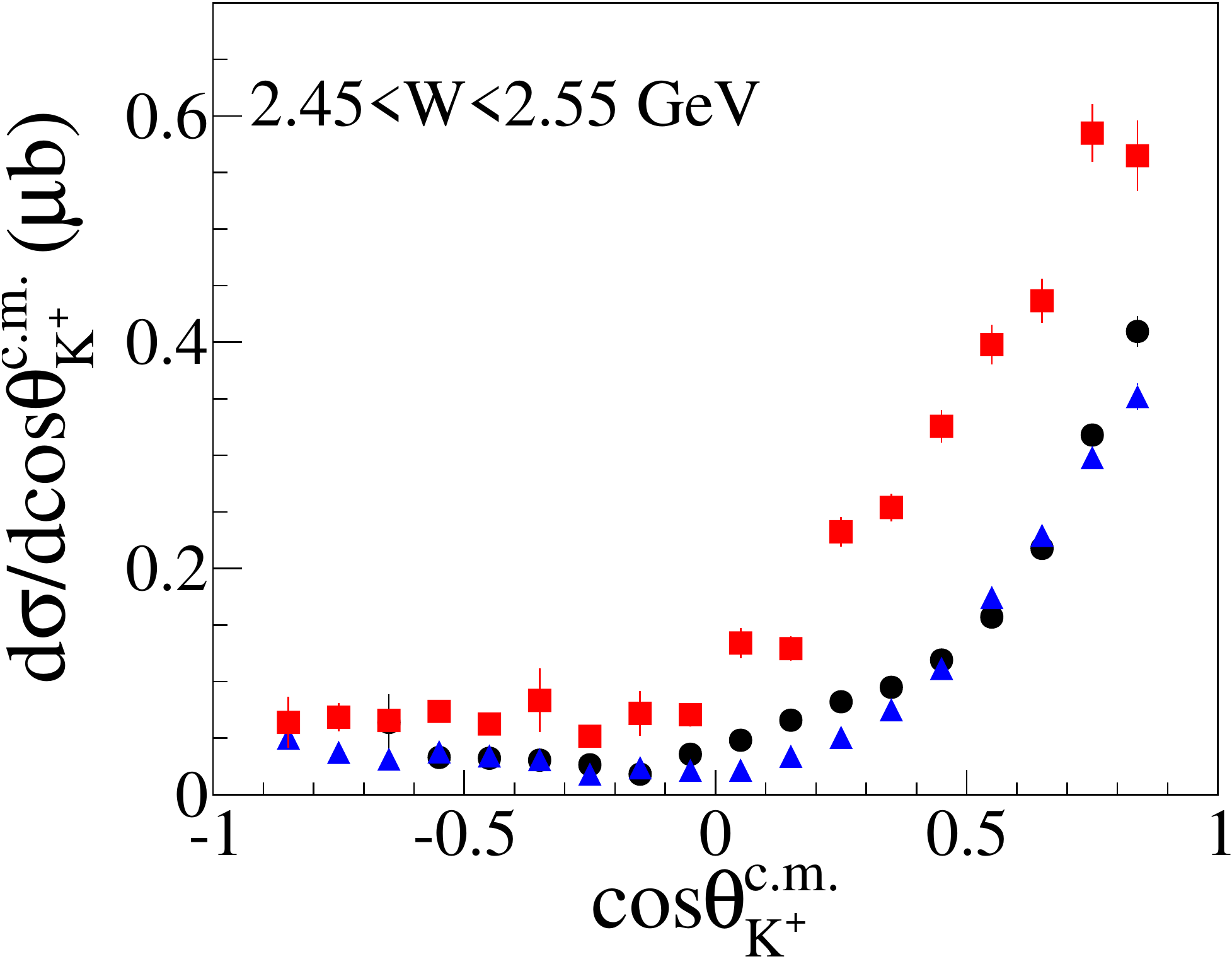}}
 \subfloat{\label{fig:xsec_cmp:W8} \includegraphics[width=0.49\textwidth]{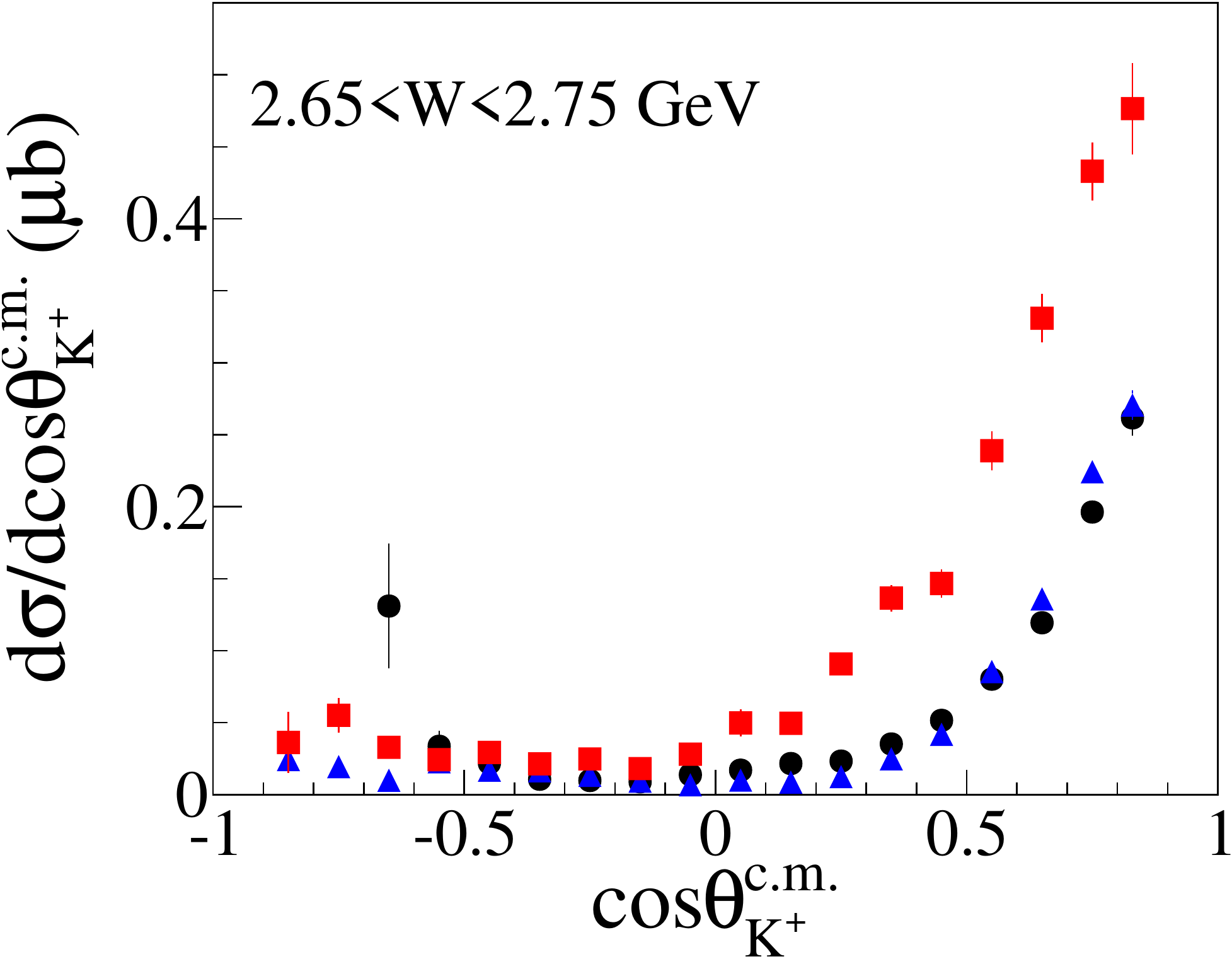}} \\
\caption{Comparison of differential cross section for $\Lambda(1405)$ (blue
  triangles), $\Sigma^{0}(1385)$ (black circles), and $\Lambda(1520)$
  (red squares) for four different $W$ bins.
\label{fig:xsec_cmp}}
\end{figure}

Next, we turn our attention to the comparison of the differential
cross sections of the three hyperons. These results present first-time
measurements of all three excited hyperons together within the same
setup.
Figures~\ref{fig:xsec_cmp}(\subref*{fig:xsec_cmp:W2}--\subref*{fig:xsec_cmp:W8})
compare
the differential cross sections of the three excited hyperons for
different energies. The $\Lambda(1405)$ is
represented by the sum of all three measured $\Sigma \pi$ channels. The
$\Sigma^{0}(1385)$ and $\Lambda(1520)$ have been corrected for the
branching ratios to $\Lambda \pi$ ($87.0\%$) and $\Sigma \pi$
($42\%$), respectively.
A comparison of the three differential cross sections shows that while
the production of $\Lambda(1520)$ is suppressed near production
threshold compared to the $\Lambda(1405)$ and $\Sigma^{0}(1385)$ due
to its heavier mass, it has a larger cross section at higher
energies.
All three hyperons exhibit forward-peaked behavior at
higher energies, suggesting that $K^{(\ast)}$-exchanges are
the main production mechanisms for all states.

\begin{figure}[h!t!pb]
\includegraphics[width=\textwidth]{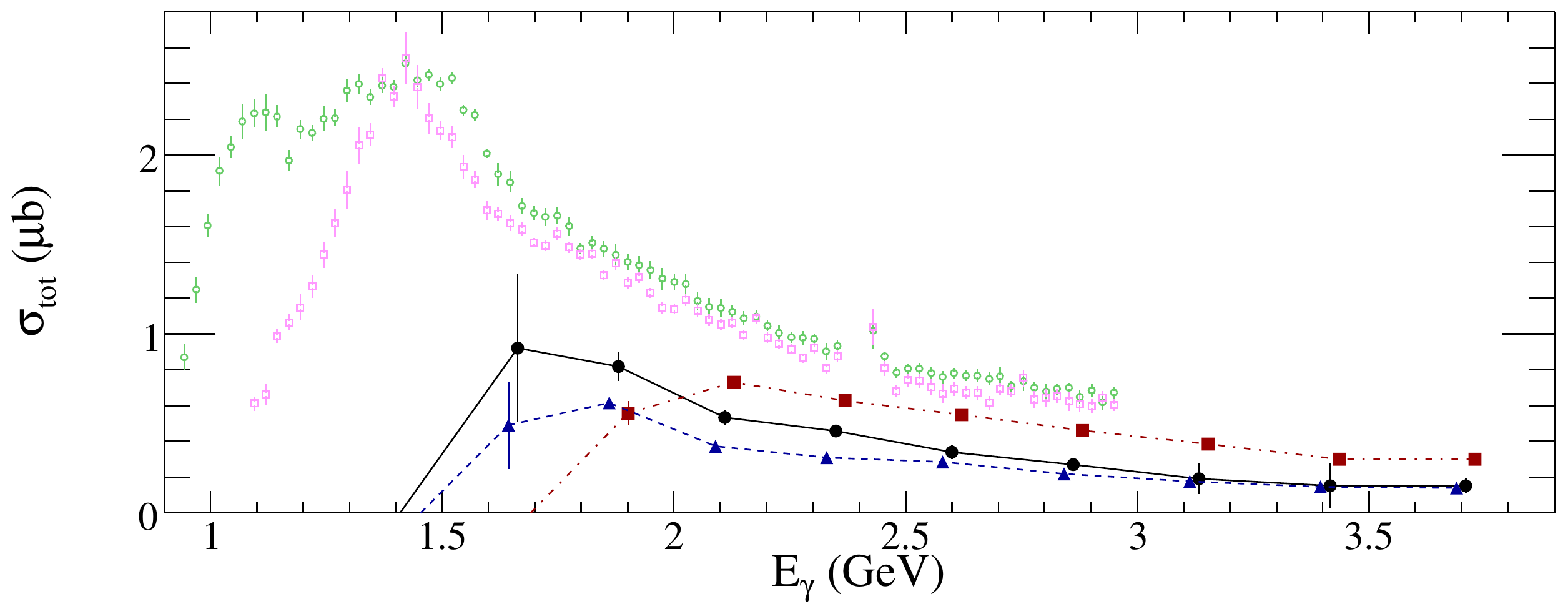}
\vspace*{8pt}
\caption{Total cross section comparisons for $\Lambda(1405)$ (blue
  triangles), $\Sigma^{0}(1385)$ (black circles), and $\Lambda(1520)$
  (red squares). For comparison, the total cross sections for $K^{+}
  \Lambda$ (open green circles) and $K^{+} \Sigma^{0}$ (open magenta
  squares) are shown.
\label{fig:sigmatot}}
\end{figure}

The differential cross sections were extrapolated to determine the
total cross sections and are shown in
Fig.~\ref{fig:sigmatot}, together
with measurements for the ground
state $K^{+} \Lambda$ and $K^{+} \Sigma^{0}$ from
Ref.~\refcite{bradford}. We see that although the $Y^{\ast}$ cross
sections are
smaller than those for the ground state hyperons, there is still
prominent production of the excited hyperons. Our binning in energy
has been driven by having enough data within each bin to measure the
$\Lambda(1405)$ line shape accurately, but further analyses of these
cross sections may show couplings to intermediate $N^{\ast}$ states. 

\section{Future Plans with $12$ GeV}

Further studies of strangeness will soon be possible at JLab, with
many upgrade efforts ongoing. The accelerator is being upgraded to a
maximum energy of $12$~GeV, and a new experimental Hall D is being
constructed, which will house the GlueX detector. The GlueX
experiment~\cite{GlueX} will be a photoproduction experiment using a
linearly polarized photon beam, with a hermetic detector for large
angular coverage. The primary goal of GlueX will be to search for
mesons with exotic quantum numbers
which are not allowed in a simple $q \overline{q}$
model. This will be done by searching for and mapping out multiplets
of mesons with specific quantum numbers, and comparing the resulting
spectrum with that of theory. Recent results from lattice
QCD~\cite{Dudek} show the existence of meson states that can be
thought as having large constituent gluon components. Comparison with
these results as well as other theories will validate whether our
understanding of how QCD works at these energies is correct.
Besides extensive searches for flavor multiplets of mesons, a rich
physics program is possible due to the all-purpose detector, and this
includes searches for excited $\Xi$ and $\Omega$ states. Installation
of detectors in the hall has already started, and beam commissioning
is scheduled to start towards the end of $2014$. With the large
statistics expected to be accumulated over years of running, there is
potential for many new and exciting studies.

\section{Summary}

The photoproduced line shapes of the $\Lambda(1405)$ in all three $\Sigma \pi$
channels have been presented, as well as the differential cross
sections of the $\Lambda(1405), \Sigma^{0}(1385)$, and
$\Lambda(1520)$. The differential cross sections of the
$\Lambda(1405)$ in each $\Sigma \pi$ channel have stark differences,
which appear in near-production-threshold energies where the
differences in line shapes are also prominent. The differential cross
sections for all three excited hyperons exhibit a strong forward peak
at higher energies, 
and the total cross sections are comparable to those of ground state
hyperon production.

\section*{Acknowledgments}

K.~M. would like to thank the organizers of the NSTAR 2013 workshop
for the invitation, hospitality, and financial support during the
workshop. Financial support from the Jefferson Science Associates
Travel Fund is gratefully acknowledged.

\end{document}